\documentclass[12pt]{iopart}

\usepackage{iopams}  
\usepackage{graphicx}
\usepackage[latin1]{inputenc} 
\usepackage[english]{babel} 
\usepackage[cyr]{aeguill}
\usepackage{amssymb}
\usepackage{textcomp}
\usepackage{bm}
\usepackage{color}
\usepackage{tocvsec2}

\newcommand{\ket}[1]{\left| #1 \right>} 

\newcommand{\rr}{\mathbf{r}}

\newcommand{\rrho}{\bm{\rho}}

\newcommand{\ex}{\mathbf{e_x}}
\newcommand{\ey}{\mathbf{e_y}}
\newcommand{\uy}{\mathbf{u_y}}
\newcommand{\ez}{\mathbf{e_z}}

\newcommand{\dd}{\mathrm{d}}

\newcommand{\mol}{{\mathrm{mol}}}

\newcommand{\hc}{{\mathrm{h.c.}}}

\newcommand{\eff}{{\mathrm{eff}}}
\newcommand{\pairs}{{\mathrm{pairs}}}
\newcommand{\Kitaev}{{\mathrm{Kitaev}}}
\newcommand{\bulk}{{\mathrm{bulk}}}

\newcommand{\mBEC}{{\mathrm{mBEC}}}
\newcommand{\SL}{{\mathrm{SL}}}

\begin{document}

\title{Realizing one-dimensional topological superfluids with ultracold atomic gases}

\author{Sylvain Nascimb\`ene}

\address{Laboratoire Kastler Brossel, CNRS, UPMC, \'Ecole Normale Sup\'erieure, 24 rue Lhomond, 75005 Paris, France}
\ead{sylvain.nascimbene@lkb.ens.fr}
\begin{abstract}
We propose  an experimental implementation of a topological superfluid with ultracold fermionic atoms. An optical superlattice is used to juxtapose a 1D gas of fermionic atoms and a 2D conventional superfluid of condensed Feshbach molecules. The latter acts as a Cooper pair reservoir and effectively induces a superfluid gap in the 1D system. Combined with a spin-dependent optical lattice along the 1D tube and laser-induced atom tunneling, we obtain a topological superfluid phase. In the regime of weak couplings to the molecular field and for a uniform gas the atomic system is equivalent to Kitaev's model of a $p$-wave superfluid. Using a numerical calculation we show that the topological superfluidity is robust beyond the perturbative limit and in the presence of a harmonic trap.
Finally we describe how to investigate some physical properties of the Majorana fermions located at the topological superfluid boundaries. In particular we discuss how to prepare and detect a given Majorana edge state. 

\end{abstract}

\maketitle

\setcounter{tocdepth}{2}
\tableofcontents

\section{Introduction}

Since the discovery of topological insulators 
\cite{hasan2010colloquium,qi2011topological}
the search for exotic states of matter with non-trivial topology such as a topological superconductor has generated strong interest. A non-trivial topological phase is characterized by a gap in the bulk elementary excitations and a non-local topological order associated with gapless states at the system edges, defects or vortex cores \cite{volovik2009universe,read2000paired,kopnin1991mutual,kitaev2001unpaired,teo2010topological,essin2011bulk}. In topological superconductors, these edge states can be described as Majorana fermions, which are quasi-particles equal to their own antiparticles \cite{majorana1937}. Besides the fundamental interest in observing Majorana fermions a great motivation for their study relies on their non-Abelian statistics, which could be used for topological quantum computation immune to decoherence from local perturbations \cite{kitaev2003fault,freedman2003topological,das2005topologically,nayak2008non,wilczek2009majorana}.

The simplest model of a topological superfluid was  introduced by Kitaev in 2001 \cite{kitaev2001unpaired}. It consists in a 1D lattice system of spinless fermions with a $p$-wave superfluid gap, described by the Hamiltonian
\begin{equation}\label{eq_HKitaev}
\hat H_\Kitaev=\sum_i-\mu\left(\hat c_i^\dagger\hat c_i^{\phantom{\dagger}}-\frac{1}{2}\right)-J\left(\hat c_i^\dagger\hat c_{i+1}^{\phantom{\dagger}}+\hc\right)+\Delta\left(\hat c_i^\dagger\hat c_{i+1}^\dagger+\hc\right),
\end{equation}
where $\hat c_i^\dagger$ creates a fermion into the lattice site $i$, and $\mu$, $J$, $\Delta$ respectively represent the chemical potential, the tunneling amplitude and the superfluid gap. However this `toy model' cannot be directly realized with solid-state systems. Subsequently several proposals described how to create topological superconductors in solid-state heterostructures combining superconducting proximity effect with a topological insulator or a semiconductor with strong spin-orbit coupling \cite{fu2008superconducting,lutchyn2010majorana,oreg2010helical,beenakker2011search,alicea2012new}. These proposals led to the first observations in 2012 of boundary quasiparticles behaving as Majorana fermions
\cite{mourik2012signatures,williams2012unconventional,rokhinson2012observation,deng2012observation}. 

Ultracold atomic gases constitute an alternative system for realizing a topological superfluid \cite{gurarie2005quantum,cooper2009stable,levinsen2011topological,stanescu2010topological,jiang2011majorana,diehl2011topology}. The extreme purity of these gases could allow one to realize model systems in a clean and controlled manner \cite{bloch2008many}. As described in \cite{kraus2012probing,liu2012probing} the study of atomic Majorana fermions would also benefit from measurement techniques specific to atomic physics. While inducing topological superfluidity using resonant $p$-wave interactions cannot be achieved due to large inelastic loss rates \cite{gurarie2005quantum,regal2003tuning,zhang2004p}, one expects that, in a quasi-2D Fermi gas, combining $s$-wave interactions with Rashba-type spin-orbit coupling may lead to the formation of a $p_x+ip_y$ superfluid, which has a non-trivial topology \cite{lin2011spin,wang2012spin,cheuk2012spin,zhang2008p_,sato2009non,gong2011bcs}. Alternatively, by analogy with proximity-induced superconductivity \cite{mcmillan1968tunneling}, superfluidity could be induced by the coherent coupling to a molecular Bose-Einstein condensate rather than by interatomic interactions \cite{jiang2011majorana}.

In this article we propose an experimental implementation of a topological superfluid with ultracold Fermi gases based on  proximity-induced superfluidity \cite{jiang2011majorana}.  We use a superlattice potential to juxtapose a quasi-2D gas of condensed Feshbach molecules and a set of 1D tubes. The coupling of atom pairs between both systems effectively creates an $s$-wave superfluid gap in the 1D tubes. In addition, the atoms in the 1D tubes are held in a spin-dependent optical lattice, in which atom tunneling is induced by the appropriate laser transitions.
The scheme makes use of several ingredients -- such as quasi-2D Fermi gases \cite{martiyanov2010observation,dyke2011crossover,frohlich2011radio,sommer2012evolution}, spin-dependent potentials \cite{mandel2003coherent,lee2007sublattice,mckay2010thermometry,soltan2011multi,karski2011direct} or laser-induced atom tunneling \cite{jaksch2003creation,aidelsburger2011experimental} -- that were already developed in the cold atom community. We show that this setup leads to the realization of a topological superfluid, which is equivalent to Kitaev's model in the perturbative regime of weak coupling to the molecular field. For stronger couplings we obtain a robust topological superfluid with a large bulk gap value, both for homogeneous and harmonically trapped systems. Finally we describe some measurable physical properties of a trapped atomic gas in this lattice geometry, which should exhibit a topological superfluid phase with Majorana fermions at its boundaries. In particular we discuss how to prepare and detect a given Majorana edge state.

\section{Proximity-induced superfluidity\label{section_1}}
Before describing the implementation of Kitaev's model, we first discuss the realization of proximity-induced $s$-wave superfluidity in a two-dimensional geometry. We consider a pair of quasi-2D atomic systems (planes $A$ and $B$) that can be created using a double-well potential along $z$ (see figure~\ref{Fig_Atom_pair_tunneling}b). The plane $B$ is filled with a two-component Fermi gas, with effective spins labelled $\ket\uparrow$ and $\ket\downarrow$. We assume that a Feshbach resonance is used to create weakly-bound Feshbach molecules \cite{frohlich2011radio,sommer2012evolution}, and that the gas is cooled sufficiently to form a nearly pure molecular Bose-Einstein condensate (mBEC). When coupled to the plane $A$ as described below, this system will act as a reservoir of Cooper pairs and effectively induce a superfluid gap in $A$ \cite{jiang2011majorana}.  

\subsection{Effective Hamiltonian of a gas coupled to a molecular BEC}

The reservoir of condensed molecules in plane $B$ is described by a coherent state with a mean number of molecules $N$. The 2D wavefunction of a single molecule is written as
\[
\phi_\mathrm{mol}(\rrho_\uparrow,\mathbf{\rrho}_\downarrow)=\frac{1}{L}\phi_{\mathrm{rel}}(\mathbf{\rrho}_\uparrow-\mathbf{\rrho}_\downarrow),
\]
where $L$ is the size of the quantification box along $x$ and $y$ and 
\[
\phi_{\mathrm{rel}}(\rrho)=\frac{1}{\sqrt{\pi}a}K_0(\rho/a)
\]
is the 2D molecular wavefunction for the relative motion. Here $a$ denotes the 2D scattering length and $K_0$ is a modified Bessel function of the second kind. The molecular binding energy will be written as $E_b=\hbar^2/ma^2$.

The coupling from the reservoir to the plane $A$ occurs via atom tunneling through the $A-B$ barrier, described by a Hamiltonian
\[
\hat H_{z}=-J_z\int\dd\rrho\;\left(\hat\psi^\dagger_{A,\uparrow}(\rrho)\hat\psi_{B,\uparrow}^{\phantom{\dagger}}(\rrho)+\hat\psi^\dagger_{A,\downarrow}(\rrho)\hat\psi_{B,\downarrow}^{\phantom{\dagger}}(\rrho)\right)+\hc,
\]
where $J_z$ is the single-atom tunneling amplitude and $\hat\psi^\dagger_{X,\sigma}(\rrho)$ denotes the creation operator of a particle of spin $\sigma$ in the plane $X$. 

We expect that tunneling of atoms towards the plane $A$ leads to the dissociation of molecules, leading to pairs of atoms of opposite spin in both planes $A$ and $B$\footnote{Note that the tunneling of bound molecules to the plane $A$ is prevented by the spin-dependent optical lattice described in the following.}. These dissociated atom pairs can be described by an effective Hamiltonian using a semi-classical approximation, which consists in replacing the mBEC by a classical field. By analogy with the Glauber unitary transform for coherent states of the electromagnetic field \cite{glauber1963coherent} we define the effective Hamiltonian as
\[
\hat H_{z,\eff}=\hat D_\mBEC^\dagger\,\hat H_{z}\,\hat D_\mBEC^{\phantom{\dagger}},
\]
where 
\[
\hat D_\mBEC^{\phantom{\dagger}}=e^{-\sqrt{N}/2}\exp(\sqrt{N}(\hat\psi_{\mol}^\dagger-\hat\psi_{\mol}^{\phantom{\dagger}}))
\]
is the displacement operator
that brings the mBEC to the vacuum state
and
\[
\psi_{\mol}^\dagger=\int\dd\rrho_\uparrow\dd\rrho_\downarrow\;\phi_\mathrm{mol}(\rrho_\uparrow,\mathbf{\rrho}_\downarrow)\hat\psi_{B,\uparrow}^\dagger(\rrho_\uparrow)\hat\psi_{B,\downarrow}^\dagger(\rrho_\downarrow)
\]
creates a molecule in plane $B$.

The effective Hamiltonian can be calculated explicitely as a series in $k_Fa$, where $k_F=2\sqrt\pi\sqrt N/L$ is the Fermi momentum in the plane $B$ (see \ref{Appendix}). As we are interested in the molecular regime $k_Fa\ll1$ we only keep for simplicity the first term in $k_Fa$, leading to a Hamiltonian  $\hat H_{z,\eff}=\hat H_z+\hat H_\pairs$, where the additional term
\[
\hat H_\pairs=-J_z\frac{k_F}{2\sqrt\pi}\int\dd\rrho_\uparrow\dd\rrho_\downarrow\;
\phi_{\mathrm{rel}}(\mathbf{\rrho}_\uparrow-\mathbf{\rrho}_\downarrow)\left(\hat\psi_{B,\uparrow}^\dagger(\rrho_\uparrow)\hat\psi_{A,\downarrow}^\dagger(\rrho_\downarrow)+\hat\psi_{A,\uparrow}^\dagger(\rrho_\uparrow)\hat\psi_{B,\downarrow}^\dagger(\rrho_\downarrow)+\hc \right)
\]
describes the creation (or annihilation) of pairs of particles dissociated from the mBEC. We show in the \ref{Appendix} that the neglected terms of the series have minor effects for the parameters used in the following. 

\begin{figure}[t!]
\includegraphics[width=\linewidth]{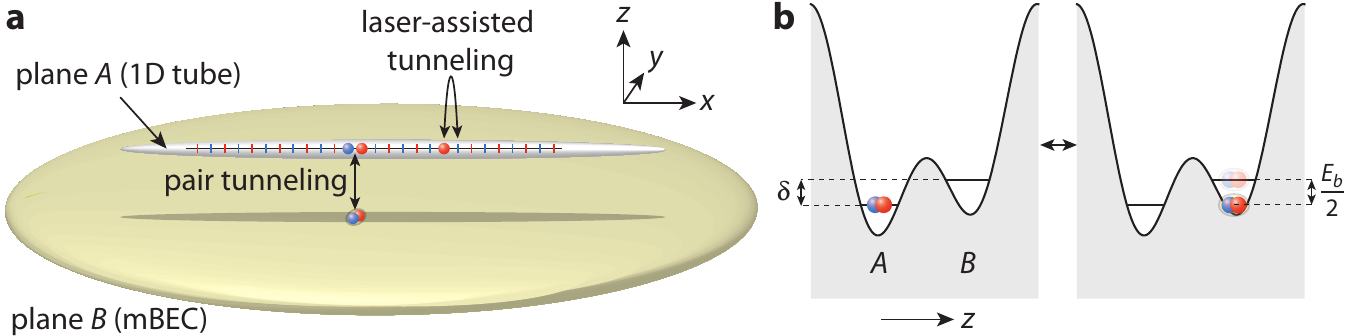}
\vspace{-5mm} \caption{
\textbf{a.} General scheme of the setup, consisting in quasi-1D gases of fermionic atoms coupled to a 2D gas of condensed molecules. The tunneling of atoms pairs into the 1D tubes acts as a superfluid gap applied externally. A spin-dependent optical lattice is applied along the tube, leading to an effective lattice
with spin $\uparrow$ ($\downarrow$) particles localized in the even (odd) sites. Atom tunneling inside the tube is driven by Raman lasers, leading to the realization of Kitaev's model of a $p$-wave superfluid. 
\textbf{b.} Scheme of the atom pair tunneling from $A$ to $B$, resonant for $E_b=2\,\delta$.
\label{Fig_Atom_pair_tunneling}}
\end{figure}

An additional energy offset $\delta$ is introduced in the plane $B$ in order to match the resonance condition $2\delta=E_b$ (see figure~\ref{Fig_Atom_pair_tunneling}b), described by the Hamiltonian
\[
\hat H_\delta=\delta\int\dd\rrho\;\left(\hat\psi^\dagger_{A,\uparrow}(\rrho)\hat\psi_{A,\uparrow}^{\phantom{\dagger}}(\rrho)+\hat\psi^\dagger_{A,\downarrow}(\rrho)\hat\psi_{A,\downarrow}^{\phantom{\dagger}}(\rrho)\right).
\]
The total effective Hamiltonian is then given by the expression 
\begin{equation}\label{eq_complete_H}
\hat H_\Delta=\hat H_z+\hat H_\pairs+\hat H_\delta.
\end{equation}

\subsection{Superfluid gap in the weak coupling regime}
A simple picture of the system can be obtained in the limit $J_z\ll\delta$ where at low energy all non-condensed atoms are located in the plane $A$ and the plane $B$ is only virtually populated by single atoms. In second-order perturbation theory, the system can be described by a Hamiltonian restricted to the plane $A$
\[
\hat H_\Delta^{(2)}=-\hat P_A\frac{(\hat H_z+\hat H_\pairs)^2}{\delta}\hat P_A,
\]
where $\hat P_A$ projects on the Hilbert space of wavefunctions localized in the plane $A$. A straightforward calculation leads to
\begin{eqnarray}\label{eq_gap}
\hat H_\Delta^{(2)}&=&\int\dd\rrho_\uparrow\dd\rrho_\downarrow\;
\Delta(\mathbf{\rrho}_\uparrow-\mathbf{\rrho}_\downarrow)\hat\psi_{A,\uparrow}^\dagger(\rrho_\uparrow)\hat\psi_{A,\downarrow}^\dagger(\rrho_\downarrow)+\hc,\\
\Delta(\rrho)&=&-\frac{k_F}{\sqrt{\pi}}\frac{J_z^2}{\delta}\phi_{\mathrm{rel}}(\mathbf{\rrho}),\nonumber
\end{eqnarray}
which describes a system of two-component fermions with an order parameter $\Delta(\rrho)$. The latter has an $s$-wave symmetry inherited from the symmetry of the molecular wavefunction. 

\section{Creating a topological superfluid\label{section_2}}
The technique of proximity-induced superfluid gap presented in section \ref{section_1} can be used to create a topological superfluid. For simplicity in this section we restrict the discussion to the limit of perturbative coupling to the mBEC for which the atom dynamics is restricted to the plane $A$.  

Optical lattices along $x$ and $y$ are applied in the plane $A$ only. The $y$ lattice freezes the motion along $y$, leading to a quasi-1D geometry. The $x$-lattice is a spin-dependent tight-binding optical lattice as described in figure~\ref{Fig_Atom_pair_tunneling}a: each spin state is held in a sublattice of period $2d$, with the two sublattices shifted by a distance $d$. This system can be viewed as a lattice of spacing $d$  with spin $\uparrow$ ($\downarrow$) particles localized in the even (odd) sites $2i$ ($2i+1$, respectively). 
The practical realization of this lattice structure is described in section \ref{section_optical_lattice}. 

The physical system restricted to the lowest band of the optical lattice can formally be viewed as a spin-polarized system of fermions on a discrete lattice. Expanding the fermionic field operators on the basis of Wannier functions:
\begin{eqnarray*}
\hat\psi_{A,\uparrow}^\dagger(\rrho)&=&\sum_iw_x(x-2id)w_y(y)\hat c_{2i}^\dagger,\\
\hat\psi_{A,\downarrow}^\dagger(\rrho)&=&\sum_iw_x(x-(2i+1)d)w_y(y)\hat c_{2i+1}^\dagger,\\
\end{eqnarray*}
we rewrite the Hamiltonian (\ref{eq_gap}) as 
\[
\hat H_\Delta^{(2)}=\sum_{i,j}\Delta_{2i,2j+1}^{(2)}\hat c_{2i}^\dagger\hat c_{2j+1}^\dagger+\hc ,
\]
where
\begin{eqnarray*}
\Delta_{2i,2j+1}^{(2)}=-\frac{k_F}{\sqrt{\pi}}\frac{J_z^2}{\delta}&\int&\dd\rrho_\uparrow\dd\rrho_\downarrow\;
\phi_{\mathrm{rel}}(\mathbf{\rrho}_\uparrow-\mathbf{\rrho}_\downarrow)\times\\&&w_y(y_\uparrow)w_y(y_\downarrow)w_x(x_\uparrow-2id)w_x(x_\downarrow-(2j+1)d).
\end{eqnarray*}
An analytic expression for $\Delta_{i,j}^{(2)}$ can be obtained assuming that the $1/e$ size $\sigma_x$ of the Wannier function along $x$ is much smaller than the lattice spacing $d$, and that the extent of the wavefunction along $y$ is much larger than the 2D scattering length $a$:
\[
\Delta_{i,j}^{(2)}=-2\sqrt{\pi}\frac{J_z^2}{\delta}k_F\sigma_xe^{-|i-j|d/a}.
\]

In the following we will use values for the scattering length $a= d/2$, for which the $\Delta_{i,j}$ is strongly suppressed except for nearest-neighbor pairs of sites, leading to the Hamiltonian
\[
\hat H_\Delta^{(2)}=\Delta^{(2)}\sum_{i}\hat c_{2i}^\dagger(\hat c_{2i+1}^\dagger+\hat c_{2i-1}^\dagger)+\hc=\Delta^{(2)}\sum_{i}(-1)^i\hat c_{i}^\dagger\hat c_{i+1}^\dagger+\hc,
\]
where 
\begin{equation}\label{eq_p-wave_gap}
\Delta^{(2)}\equiv\Delta_{i,i+1}^{(2)}=-2\sqrt{\pi}\frac{J_z^2}{\delta}k_F\sigma_xe^{-d/a}.
\end{equation}
This Hamiltonian differs from the gap term in Kitaev's model due to the $(-1)^i$ factors. However, the latter can be eliminated by redefining the operators $\hat c_i$ according to $\hat c_{2i}\rightarrow (-1)^i\hat d_{2i}$ and $\hat c_{2i+1}\rightarrow (-1)^i\hat d_{2i+1}$. 

The complete Kitaev model 
\[
\hat H_\Kitaev=\sum_i-\mu\left(\hat d_i^\dagger\hat d_i^{\phantom{\dagger}}-\frac{1}{2}\right)-J\left(\hat d_i^\dagger\hat d_{i+1}^{\phantom{\dagger}}+\hc\right)+\Delta\left(\hat d_i^\dagger\hat d_{i+1}^\dagger+\hc\right)
\]
is then obtained by introducing a small additional energy  offset $\mu$ of the plane $A$ and inducing the required atom tunneling $J$ between neighboring sites. The latter is performed using Raman transitions that flip the atom spin, which induces atom tunneling with the required phase $(-1)^i$ of the hopping operator $\hat d_i^\dagger\hat d_{i+1}=(-1)^i\hat c_i^\dagger\hat c_{i+1}$ expressed with the physical operators $\hat c_i$ \cite{jaksch2003creation}. The required laser setup is described in section~\ref{section_optical_lattice}.

\section{Bogoliubov spectra and Majorana fermions\label{section_numerical}}
The topological character of a topological superfluid can be characterized by its spectrum of Bogoliubov excitations. Let us remind the main properties of Bogoliubov excitations in the case of Kitaev's model, described by the Hamiltonian (\ref{eq_HKitaev}) \cite{kitaev2001unpaired}. In the case of a homogeneous 1D chain with periodic boundary conditions, the spectrum of Bogoliubov excitations is gapped when $\Delta>0$, except when the chemical potential is fine-tuned to $\mu=-2J$ or $2J$. These points indicate phase transitions between topologically different systems. The phase $\mu<-2J$ can be smoothly connected to the trivial vacuum by decreasing $\mu$ to $-\infty$ without closing the gap, it is thus non-topological. The phase $\mu>2J$ is essentially equivalent to $\mu<-2J$ since both phases are related by a particle-hole transform. On the other hand, the phase $-2J<\mu<2J$ has a non-trivial topology which can be characterized by a $\mathbb{Z}_2$ topological invariant \cite{kitaev2001unpaired}.

\begin{figure}[t!]
\begin{center}
\includegraphics[width=0.8\linewidth]{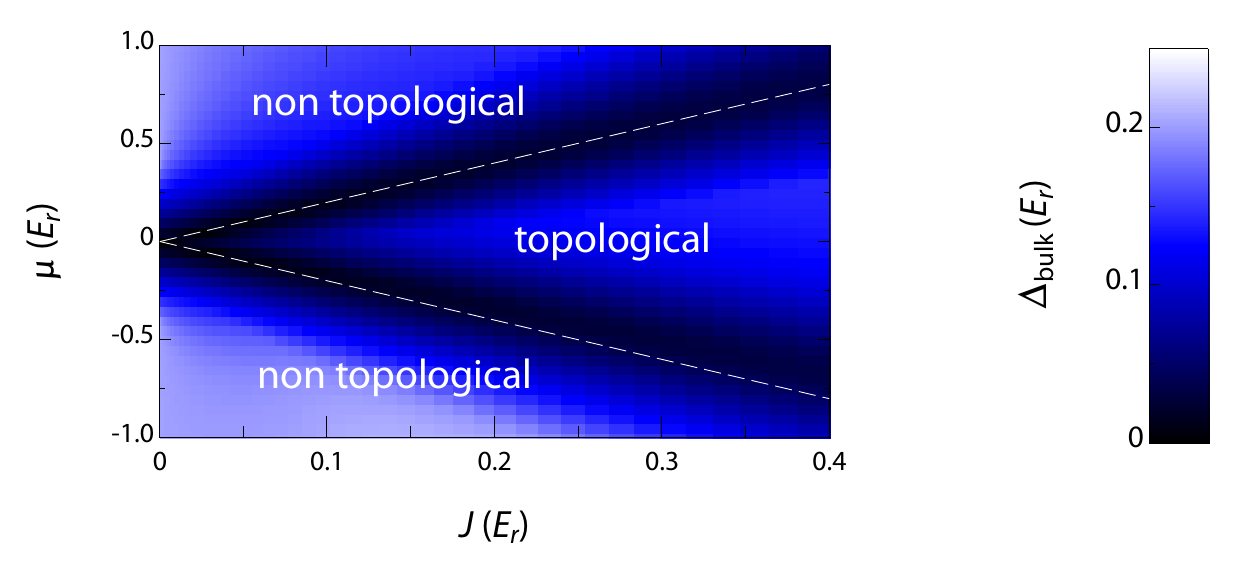}
\end{center}
\vspace{-7mm} \caption{
Amplitude of the bulk superfluid gap $\Delta_\bulk$  as a function of the tunnel coupling $J$ and the chemical potential $\mu$ calculated for a system of 30 lattice sites with periodic boundary conditions. The topological transition corresponds to the lines for which the gap cancels; their location agrees well with the prediction of Kitaev's model $\mu=\pm2J$ (dashed white lines).
 \label{Fig_Phase_diagram}}
\end{figure}

Similarly to edge states in the topological quantum Hall phases, for a chain with open boundary conditions the non-trivial topology is associated with boundary states. In the case of 1D topological superfluids the boundary states form a pair of Majorana modes occuring at zero energy, i.e. in the middle of the gap of bulk Bogoliubov excitations. These states are highly non-local -- delocalized at both ends of the topological superfluid -- and become exponentially degenerate when increasing the size of the topological phase.

We present in this section spectra of Bogoliubov excitations calculated for realistic experimental parameters, using the Hamiltonian described in sections \ref{section_1} and \ref{section_2} (without performing the perturbation theory to second order in $J_z/\delta$). The system parameters are the following: the lattice spacing $d$ is equal to half the wavelength $\lambda$ of the optical lattice (see section~\ref{section_optical_lattice}), and we choose a 2D scattering length value $a=d/2$ so that the proximity-induced gap is restricted to nearest neighbours. This scattering length value corresponds to a molecular binding energy $E_b\simeq 0.8\,E_r$, where $E_r=h^2/2m\lambda^2$ denotes the recoil energy. The molecule density in $B$ is fixed to a value $n_\mol=0.16\,d^{-2}$ corresponding to a Fermi energy $E_F=E_b/4$, i.e. on the BEC side of the BEC-BCS crossover of $s$-wave Fermi superfluids. This condition ensures that at low temperature the condensed fraction of the mBEC tends to unity \cite{salasnich2007condensate}. We choose an $x$-lattice depth value $V_x=5\,E_r$, which leads to a spontaneous atom tunneling of negligible amplitude $6.2\times10^{-4}\,E_r$. The band gap is then equal to $2.0\,E_r$, a value  larger than the energy offset $\delta=E_b/2=0.4\,E_r$, as required by the single-band assumption made above.

For the Raman-induced tunneling rate we take a value on the order of the $p$-wave gap $\Delta^{(2)}$ calculated in second-order perturbation theory, which should maximize the system bulk gap $\Delta_\bulk$ in the perturbative limit $J_z\ll\delta$ (more precisely $\Delta_\bulk=2\,\Delta$ for $J=\Delta$ for Kitaev's model \cite{kitaev2001unpaired}).

\subsection{Homogeneous topological superfluid}
We discuss in this section the case of a homogeneous system.

We first calculate the spectrum of Bogoliubov excitations for a homogeneous system of 30 lattice sites with periodic boundary conditions. It exhibits a bulk gap whose amplitude $\Delta_\bulk$  is plotted in figure~\ref{Fig_Phase_diagram} as a function of the tunnel coupling $J$ and the chemical potential $\mu$. We observe a cancellation of the bulk gap on two lines of the phase diagram $\mu\simeq\pm2J$, in agreement with the prediction of Kitaev's model. The lines indicate the location of the topological transition.

\begin{figure}[t!]
\includegraphics[width=\linewidth]{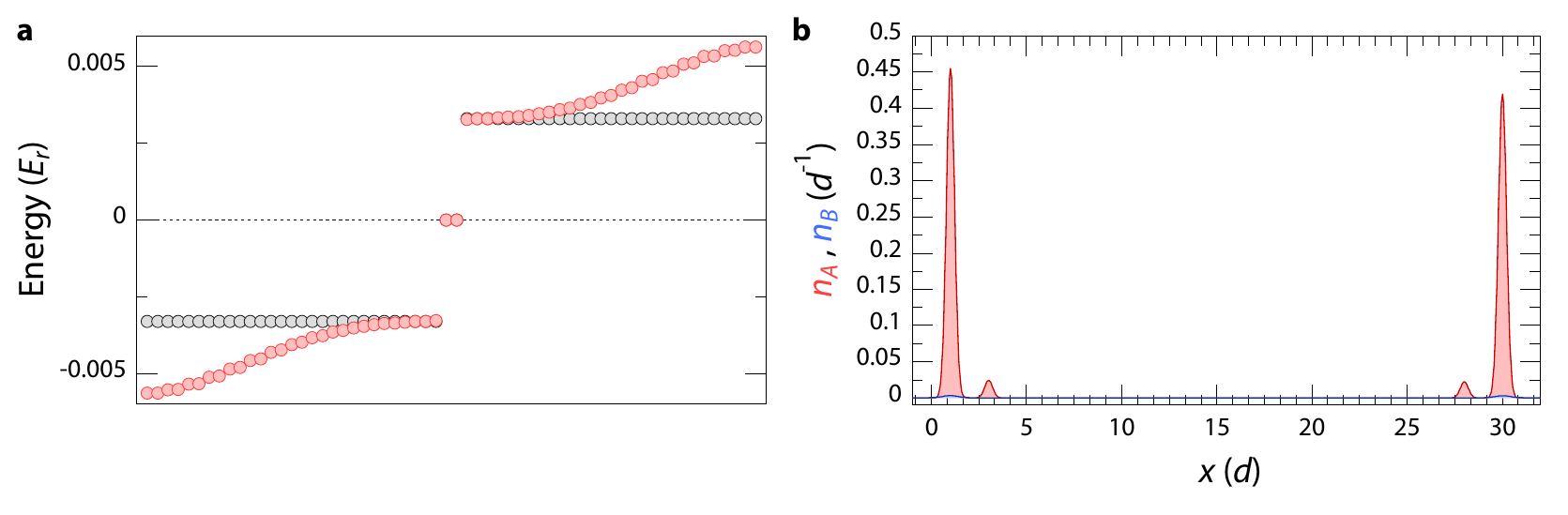}
\vspace{-10mm} \caption{
\textbf{a.} Spectrum of Bogoliubov excitations (red dots) calculated for  $J_z=0.2\,\delta$ and $J=\Delta^{(2)}$, and compared with the prediction of Kitaev's model with $J=\Delta=\Delta^{(2)}$ (black dots). 
\textbf{b.} Density distribution along $x$ of a zero-energy Majorana state, in the planes $A$ (red line) and $B$ (blue line), revealing the non-local character of Majorana states. In the perturbative regime $J\ll\delta$ the population in $B$ remains small. 
 \label{Fig_Homogeneous_weak_coupling}}
\end{figure}

We then repeat the calculation of Bogoliubov spectra for homogeneous systems with sharp boundaries, for which zero-energy Majorana end states are expected in the topological phase. We first consider a small tunneling rate value $J_z=0.2\,\delta$. As shown in figure~\ref{Fig_Homogeneous_weak_coupling}a, the calculated spectrum of Bogoliubov excitations is very similar to the one of Kitaev's lattice model, with a bulk gap and two Majorana states at zero energy. The full excitation spectrum quantitatively differs from Kitaev's model due to additional effects such as the creation of atom pairs in non-neighbouring lattice sites; yet the bulk gap value is in good agreement with the second-order perturbation result of equation~(\ref{eq_p-wave_gap}).

\begin{figure}[t!]
\includegraphics[width=\linewidth]{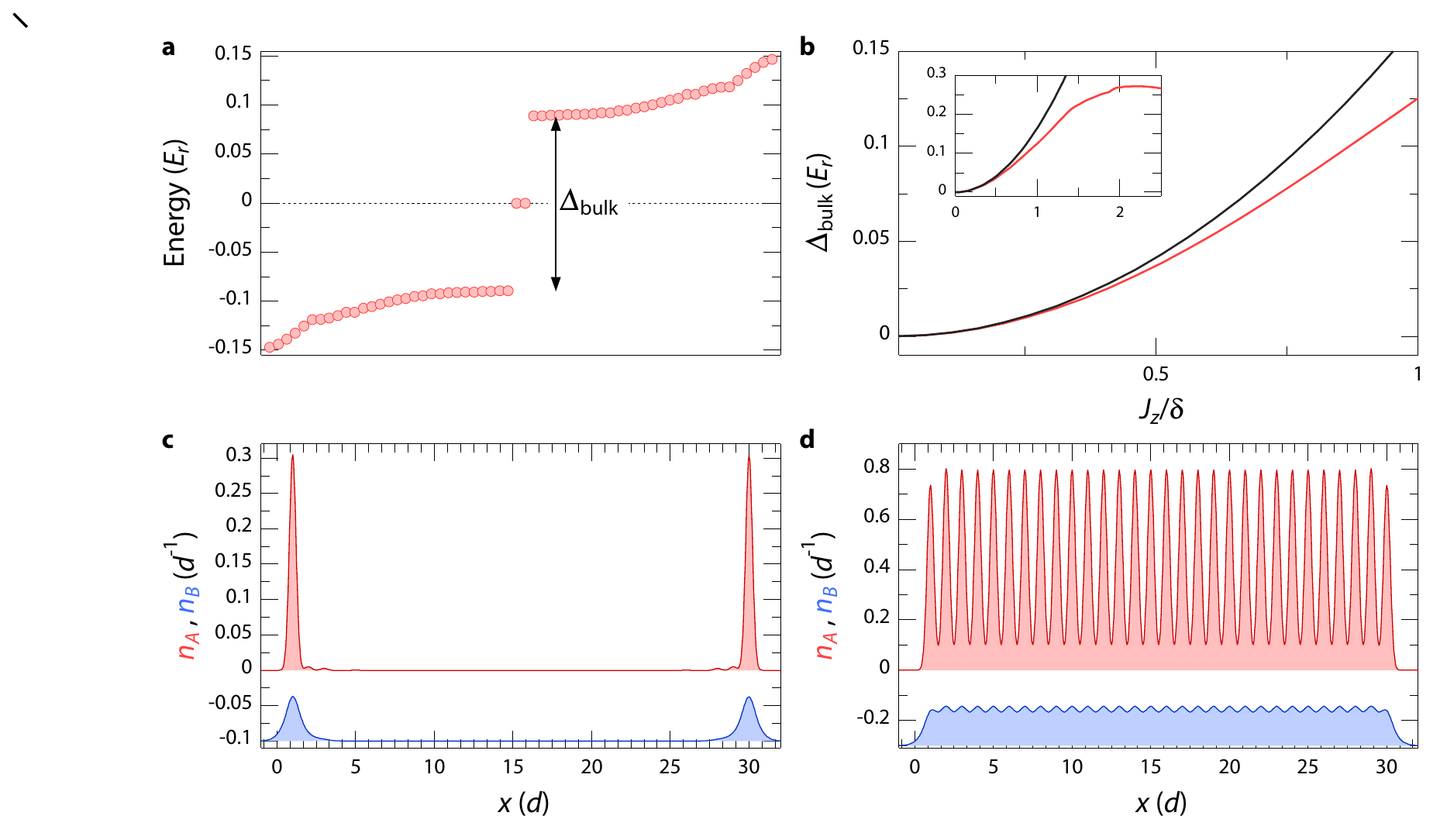}
\vspace{-8mm} \caption{
\textbf{a.} Spectrum of Bogoliubov excitations (red dots) for a homogeneous system with sharp boundaries, calculated for  $J_z=\delta$, and $J=2\,\Delta^{(2)}$. It exhibits a bulk gap $\Delta_\bulk=0.18\,E_r$ and a pair of zero-energy Majorana states with a residual splitting $\Delta_s\sim10^{-12}\,E_r$. 
\textbf{b.} Evolution of the bulk gap amplitude $\Delta_\bulk$ as a function of the ratio $J_z/\delta$ (red line), for $J=\Delta^{(2)}$, with $\Delta^{(2)}$ given by eq.\,(\ref{eq_p-wave_gap}). The black line represents the prediction of second-order perturbation theory $\Delta_\bulk=2\,\Delta^{(2)}$.
\textbf{c.} Density distribution along $x$ of a zero-energy Majorana state, in the planes $A$ (red line) and $B$ (blue line, offset for clarity). In the strong coupling regime $J\sim\delta$ the population in $B$ is not negligible. 
\textbf{d.} Total density distribution along $x$ calculated at zero temperature. Majorana states are not visible in this almost uniform density profile. \label{Fig_Homogeneous_strong_coupling}}
\end{figure}

When increasing the tunneling rate value beyond the weak coupling regime $J_z\ll\delta$, we find that the structure of the Bogoliubov spectra qualitatively remains the same for tunneling rates $J_z\sim\delta$ (see figure~\ref{Fig_Homogeneous_strong_coupling}a). This should allow one to create $p$-wave superfluids with large bulk gap values $\Delta_\bulk\sim0.2\,E_r$, which corresponds to a gap of $k_B\times40\,$nK for the physical system described in section \ref{description} (see figure~\ref{Fig_Homogeneous_strong_coupling}b). As the gas temperature has to be much smaller than the Fermi energy $E_F\simeq0.2\,E_r$ in order to create a molecular BEC in plane $B$ \cite{petrov2003superfluid}, we expect the $p$-wave superfluid to be essentially in the zero-temperature regime for $J_z\sim\delta$. 

The density probability of the Majorana states at zero energy is shown in figure~\ref{Fig_Homogeneous_strong_coupling}c in the case $J_z=\delta$. In the plane $A$ essentially only the lattice sites at the edge of the system are populated. We note that in that case the probability to occupy the plane $B$ is not negligible (about 30$\%$). The total atom density probability is almost uniform and close to half filling (see figure~\ref{Fig_Homogeneous_strong_coupling}d).

\subsection{Topological superfluid in a harmonic trap}
Homogeneous systems with sharp boundaries are difficult to realize with ultracold atom systems. We thus now focus on gases held in a harmonic trap $V(x)=\frac{1}{2}m\omega^2x^2$. In a local density approximation picture, the local properties of the gas are expected to be close to the ones of a homogeneous gas with a local chemical potential $\mu-V(x)$. A gas of chemical potential $-2J<\mu<2J$ should thus exhibit a phase separation with a topological superfluid phase in the trap bottom ($\mu-V(x)>-2J$) and trivial superfluids at the edges ($\mu-V(x)<-2J$). We also expect the presence of Majorana zero-energy states at the boundary of the topological superfluid. 

\begin{figure}[t!]
\includegraphics[width=\linewidth]{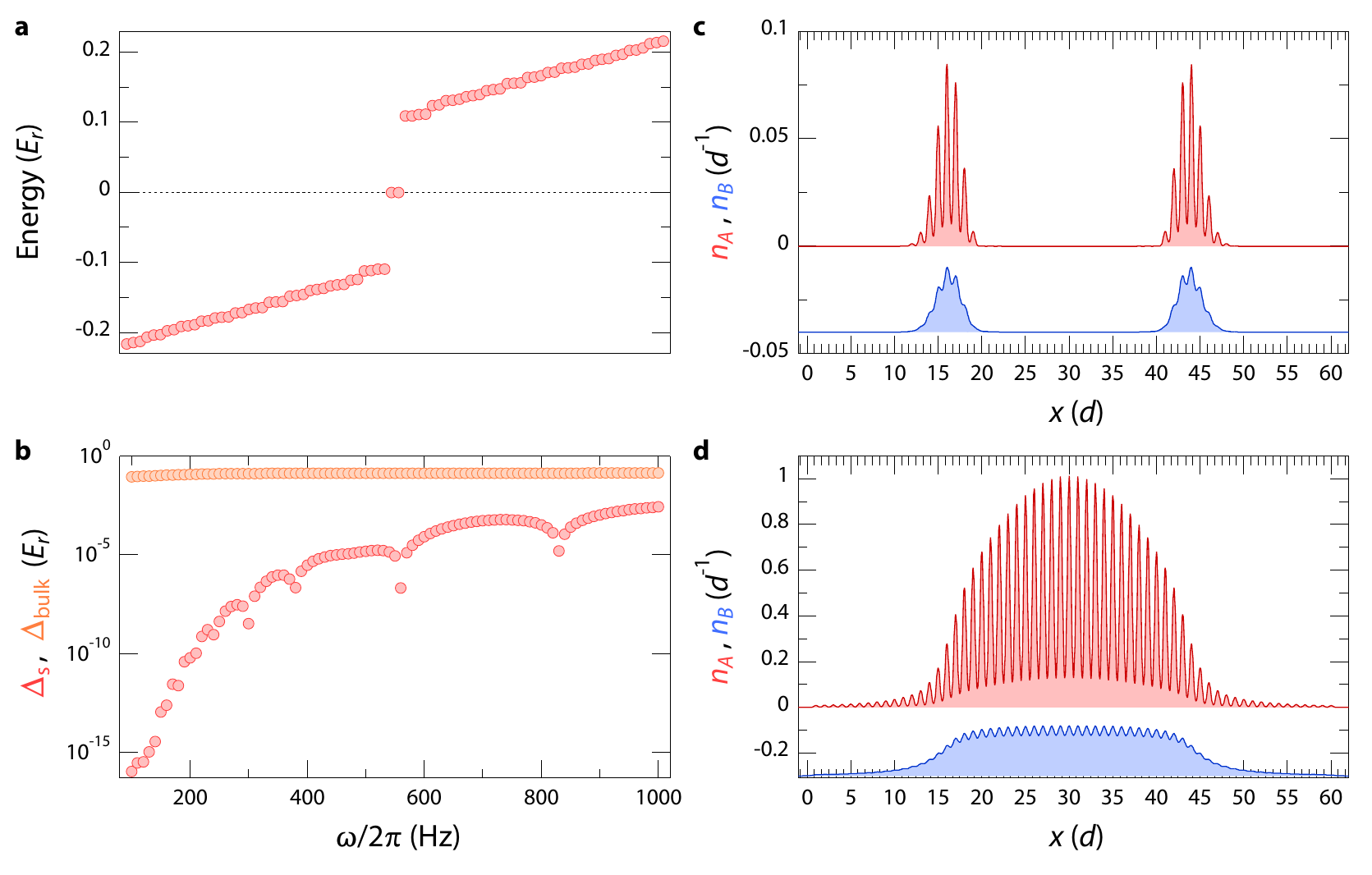}
\vspace{-8mm} \caption{
\textbf{a.} Spectrum of Bogoliubov excitations (red dots) for a harmonically trapped system, calculated for  $J_z=1.5\,\delta$, $J=2\,\Delta^{(2)}$, $\omega=2\pi\times200$\,Hz. It exhibits a bulk gap $\Delta_\bulk=0.22\,E_r$ and a pair of zero-energy Majorana states with a residual splitting $\Delta_s\sim10^{-10}\,E_r$. 
\textbf{b.} Evolution of the energy splitting $\Delta_s$ between Majorana states (red dots) and of the bulk gap $\Delta_\bulk$ (orange dots) as a function fo the trapping frequency $\omega$. While the bulk gap essentially does not depend on $\omega$, the splitting between Majorana states strongly decreases when decreasing the $\omega$ value, i.e. increasing the separation between Majorana fermions.
\textbf{c.} Density distribution along $x$ of a zero-energy Majorana state, in the planes $A$ (red line) and $B$ (blue line). \textbf{d.} Total density distribution along $x$ calculated at zero temperature.  \label{Fig_Harmonic_strong_coupling}}
\end{figure}

For the parameters given in the caption of figure~\ref{Fig_Harmonic_strong_coupling}, the calculated spectrum of Bogoliubov excitations is  gapped with a pair of quasi-degenerate Majorana end states at zero energy (see figure~\ref{Fig_Harmonic_strong_coupling}a). These states are non-local, being the superposition of two localized wavefunctions ($1/e$ size of $\simeq2.5\,d$) separated by $\simeq21\,d$ (see figure~\ref{Fig_Harmonic_strong_coupling}c). The position of these zero-energy modes agrees well with the position of the topological transition given by the local density approximation. Note that the Majorana states are not directly visible in the total density profile; yet at the location of the topological transition the gas compressibility sharply varies (see figure~\ref{Fig_Harmonic_strong_coupling}d), which could be useful to establish the phase diagram of the system \cite{liu2012topological,seo2012topological}.

 A figure of merit of the topological protection can be given by the residual splitting between the two Majorana states. As shown in figure~\ref{Fig_Harmonic_strong_coupling}b, the topological protection improves by increasing the separation between the Majorana fermions, which can be performed by opening the trap along $x$.

\subsection{Probing the Majorana fermions}
As shown in figure~\ref{Fig_Harmonic_strong_coupling}c,d, the presence of Majorana end states cannot be revealed from the total atom density. Photo-emission spectroscopy \cite{stewart2008using}, which directly addresses the structure of the Bogoliubov excitation spectrum, is a convenient probe for these states \cite{kraus2012probing,liu2012probing}. It consists here in coherently coupling the spin $\ket\uparrow$ atoms  to an unpopulated internal state $\ket\alpha$, which can be probed either in situ or after time-of-flight. The photoemission performed at a given frequency $\Omega$ gives rise to a transfer of atoms occupying the Bogoliubov excitations of energy $E$ matching the resonance condition $\hbar\Omega=\hbar\omega_0-E$, where $\omega_0$ is the resonance frequency for a single atom. The energy of the final state $\ket\alpha$ is neglected for simplicity.

\subsubsection{Density probability of Majorana states.}

\begin{figure}[t!]
\begin{center}
\includegraphics[width=0.7\linewidth]{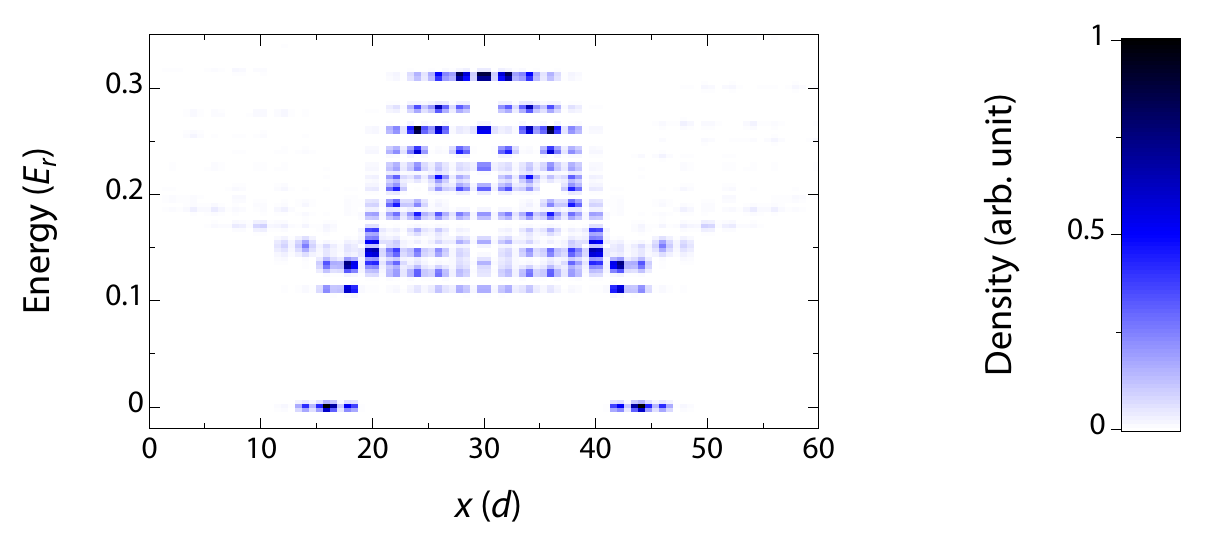}
\end{center}
\vspace{-8mm} \caption{Position-resolved photoemission spectrum calculated with the parameters of figure~\ref{Fig_Harmonic_strong_coupling}. The Majorana states probed at zero energy are localized at the edges of the topological superfluid.
\label{Fig_photoemission_in_situ}}
\end{figure}

In order to probe the density probability of Majorana states, we probe the target state density distribution right after the photoemission pulse. 
The target spin state $\ket\alpha$ can be chosen as another Zeeman state that would be trapped in $x$ lattice, with a photoemission coupling provided by a radiofrequency field. We assume that the photoemission spectroscopy is performed in the linear perturbative regime, and that the target spin state $\ket\alpha$ can be imaged in situ with single-atom sensitivity \cite{bakr2009quantum,sherson2010single}. The density probability of $\ket\alpha$ atoms as a function of the driving frequency is shown in figure~\ref{Fig_photoemission_in_situ}. The bulk Bogoliubov excitations are associated with a broad energy spectrum above the superfluid gap. The Majorana boundary states correspond to a narrow feature at zero energy, with photoemitted atoms  located at both ends of the topological superfluid phase.

\subsubsection{Momentum distribution of Majorana states.}
The photoemission technique can also be applied to extract the momentum distribution of Majorana states, by imaging the target state $\ket{\alpha}$ after time-of-flight. 

For this measurement the target state $\ket\alpha$ should be untrapped so that its eigenstates have a well-defined momentum. A possible choice for $\ket\alpha$ is a long-lived excited state\footnote{such as the $4f^{9}(^{6}H^{0})5d6s^{2}\;^{7}H^{0}_8$ metastable state of Dy ($1/e$ lifetime of 7 ms), coupled to the ground electronic state by an optical transition at 1322\,nm \cite{dzuba2010theoretical}.}, for which the optical dipole force of wavelength $\lambda$ used for the optical lattice would be very weak \cite{dzuba2011dynamic}. The photoemission coupling would be achieved using the ultranarrow optical transition between the ground electronic state and the state $\ket\alpha$.  

\begin{figure}[t!]
\begin{center}
\includegraphics[width=\linewidth]{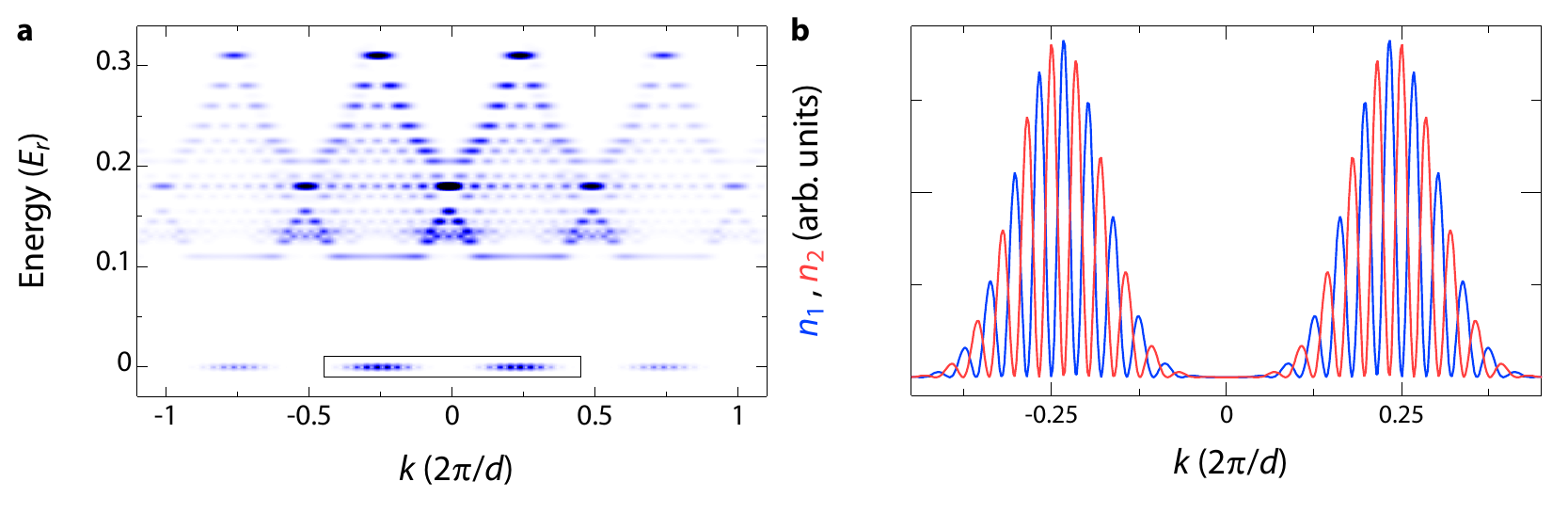}
\end{center}
\vspace{-7mm} \caption{
\textbf{a.} Momentum-resolved photoemission spectrum, exhibiting an oscillatory behavior at zero energy reflecting the delocalization of Majorana states at both ends of the topological superfluid. The small rectangle at zero energy represents the spectrum of Majorana states shown in \textbf{b}.
\textbf{b.} Momentum-resolved photoemission spectrum $n_i(k)$ for both Majorana states $i=1,2$, differing from the phase of the fast oscillations. The slow modulation (period $\pi/d$) reflects the wavefunction modulation by the lattice potential, while the fast oscillation is due to the delocalization of the wavefunction at both ends of the topological superfluid. \label{Fig_photoemission_TOF}}
\end{figure}

The expected photoemission spectrum (shown  in figure~\ref{Fig_photoemission_TOF}a) can be divided in a broad spectrum of bulk excitations and a narrow feature at zero energy corresponding to the Majorana states. The oscillatory behavior at zero energy reflects the quantum superposition of the wavefunction of Majorana states at both ends of the topological superfluid. Besides demonstrating the non-local character of the Majorana states, this probe also allows one to discriminate between the two zero-energy states, whose phase of the interference pattern differ (shown  in figure~\ref{Fig_photoemission_TOF}b).

\subsubsection{Adiabatic preparation of a given Majorana state.}
In the previous section we described how to discriminate between Majorana states using momentum-resolved photoemission spectroscopy. This raises the question of the preparation of the system in a given Majorana state \cite{kraus2012probing}, which is an important prerequisite to the manipulation of Majorana fermions for quantum information purposes. We propose to prepare a given Majorana state by crossing adiabatically a phase transition from a trivial to a topological superfluid phase. This can be achieved by changing adiabatically the chemical potential $\mu$, which can be performed by dynamically varying the double-well energy offset (see section \ref{section_superlattice}). 

\begin{figure}[t!]
\begin{center}
\includegraphics[width=\linewidth]{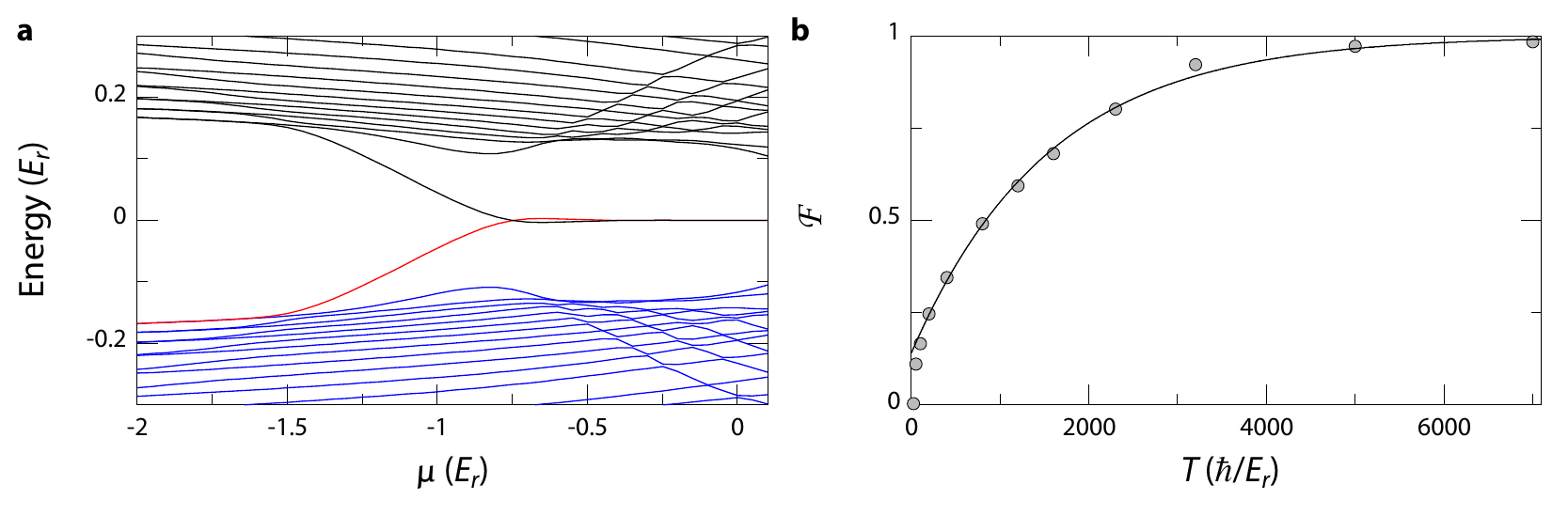}
\end{center}
\vspace{-7mm} \caption{
\textbf{a.} Energy of Bogoliubov excitations as a function of the chemical potential $\mu$. The parameters are the same than for figure~\ref{Fig_Harmonic_strong_coupling}, expect with a trap frequency $\omega/2\pi=600\,$Hz. The bulk hole sector is pictured in blue. Starting with a system prepared at low temperature in the trivial phase ($\mu\simeq-2\,E_r$) the Majorana state connected to the hole sector of the trivial superfluid (in red) can be prepared by increasing adiabatically the chemical potential up to $\mu=0$.
\textbf{b.} Expected fidelity $\mathcal{F}$ as a function of the ramp duration $T$ (black dots). The solid line is an exponential fit to the numerical data. A fidelity $\mathcal{F}>0.9$ is expected for ramp durations $T>3300\,\hbar/E_r\simeq120\,$ms.  \label{Fig_Adiabatic_Ramp}}
\end{figure}

In the local density approximation framework, one expects a topological superfluid phase to nucleate in the trap bottom when the chemical potential corresponds to the critical value for topological superfluidity $\mu_c$ at the bottom of the trap ($\mu_c\simeq-2\,J$). As shown in figure~\ref{Fig_Adiabatic_Ramp}a, the nucleation of the topological superfluid is associated with the creation of Majorana zero-energy states  from two bulk states (one from the hole sector, one from the particle sector) when increasing the chemical potential value across $\mu=\mu_c$. By increasing adiabatically the chemical potential value from a trivial to a topological phase, one thus expects the system to populate the Majorana state which is adiabatically connected to the hole sector of the trivial superfluid. In order to test the feasability of this scheme, we simulated the effect of the finite duration of the chemical potential ramp by solving the Schrödinger equation for our system. As shown in  figure~\ref{Fig_Adiabatic_Ramp}b, fidelities $\mathcal{F}>0.9$ can be achieved with the parameters described in the caption of figure~\ref{Fig_Adiabatic_Ramp} for ramp durations of about $120\,$ms. Note that shorter adiabatic paths could be used if one crosses the topological transition with a lower atom number in plane $A$ (i.e. higher trapping frequency $\omega$), and then increase the atom number in $A$ by opening the trap far from the topological transition so that Majorana states are deeply topologically protected.

\section{Description of the experimental setup\label{description}}

\subsection{Choice of the atomic species}
The proposal described above makes uses of a spin-dependent optical lattice potential. For fermionic alkali atoms ($^6$Li and $^{40}$K) spin-dependent optical dipole potentials are associated with large heating rates due to spontaneous emission \cite{wang2012spin,cheuk2012spin}, because of the small fine-structure splitting of the $P$ electronic levels compared to the transition linewidth. In order to circumvent this issue we propose to use a fermionic atom of the Lanthanide family, such as $^{167}$Er or $^{161}$Dy \cite{lu2011strongly,lu2012quantum,aikawa2012bose}. 
These atoms benefit from narrow optical transitions that can be used to create spin-dependent dipole potentials with small heating rates. For example, strong spin-dependent potentials can be created for Dy atoms using optical lattices at $\lambda=530$\;nm, close to the 530.3\;nm transition 
 (electronic angular momenta $J=8\rightarrow J'=7$, linewidth $\Gamma=2\pi\times184\,$kHz).

Due to the large magnetic moment of these atoms, we expect large dipolar relaxation rates, except for mixtures of the two Zeeman states of lowest energy in the case of a fermionic species \cite{pasquiou2010control}. As shown in the case of $^{168}$Er Bose-Einstein condensates, the very large number of Zeeman states in the electronic ground state should lead to numerous Feshbach resonances, making it possible to create a molecular Bose-Einstein condensate for large scattering length values \cite{aikawa2012bose}.

\subsection{Optical lattice configuration\label{section_optical_lattice}}
We describe in this section the laser configuration required to create the double-well potential along $z$, the plane-dependent lattices along $x$ and $y$, and the Raman lasers driving atom tunneling in the $x$ lattice.

\subsubsection{Superlattice potential along $z$\label{section_superlattice}.}
\begin{figure}[t!]
\begin{center}
\includegraphics[width=0.8\linewidth]{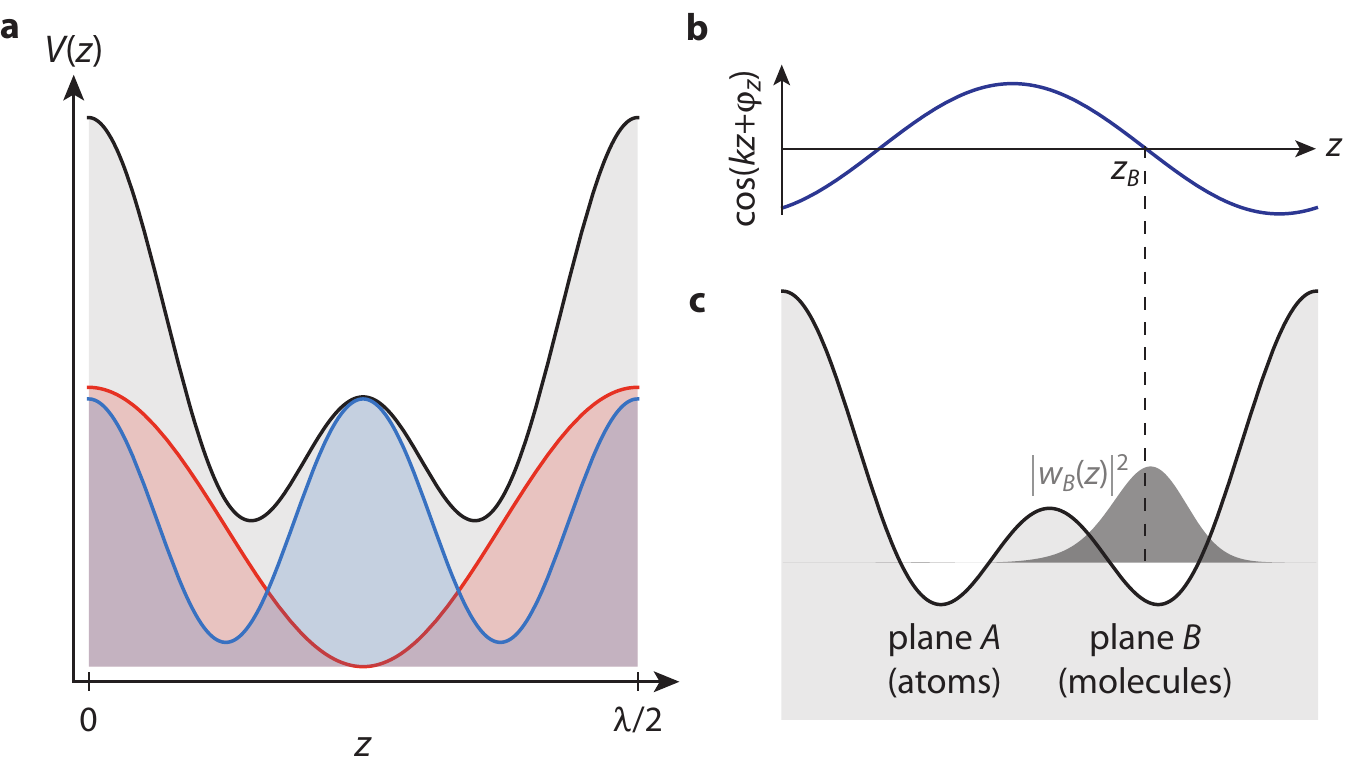}
\end{center}
\vspace{-6mm} \caption{
\textbf{a.} Scheme of the bichromatic superlattice potential along $z$ (black line), made of the superposition of two standing waves of wavelength $\lambda$ (blue line) and $2\,\lambda$ (red line). The energy offset $\delta$ in the resulting double-well potential can be controlled using a relative offset in position of the two standing waves.
\textbf{b.} Amplitude of the standing wave along $z$ used for generating the $y$ lattice, which vanishes at $z_B$. 
\textbf{c.} Superlattice potential with the Wannier function in plane $B$ centered on $z_B$. 
 \label{Fig_Lattice}}
\end{figure}

The atoms are confined in a bilayer system of two planes $A$ and $B$ using a double-well potential. The latter can be created using a bichromatic optical lattice along $z$, i.e. the superposition of two standing waves of wavelength $\lambda$ and $2\lambda$, leading to a potential
\begin{equation}\label{eq_VSL}
V_\SL(z)=V_1\cos^2(kz)+V_2\cos^2(kz/2+\phi),
\end{equation}
where $k=2\pi/\lambda$ \cite{folling2007direct} (see figure~\ref{Fig_Lattice}a).
Atom tunneling towards other planes is suppressed using lattice depths of about $20\,E_r$. The lattice intensities $V_1$, $V_2$ and relative phase $\phi$ enable one to control independently the tunnel coupling $J_z$ between $A$ and $B$ and the value of the energy offsets $\delta$ and $\mu$.

\subsubsection{Plane-dependent lattices along $x$ and $y$\label{section_lattice_y}.}
An optical lattice is used to freeze the atom motion along $y$ in plane $A$. The atom motion in plane $B$, which contains the two-dimensional mBEC, should remain unaffected so that phase coherence in the mBEC is maintained.

\begin{figure}[t!]
\includegraphics[width=\linewidth]{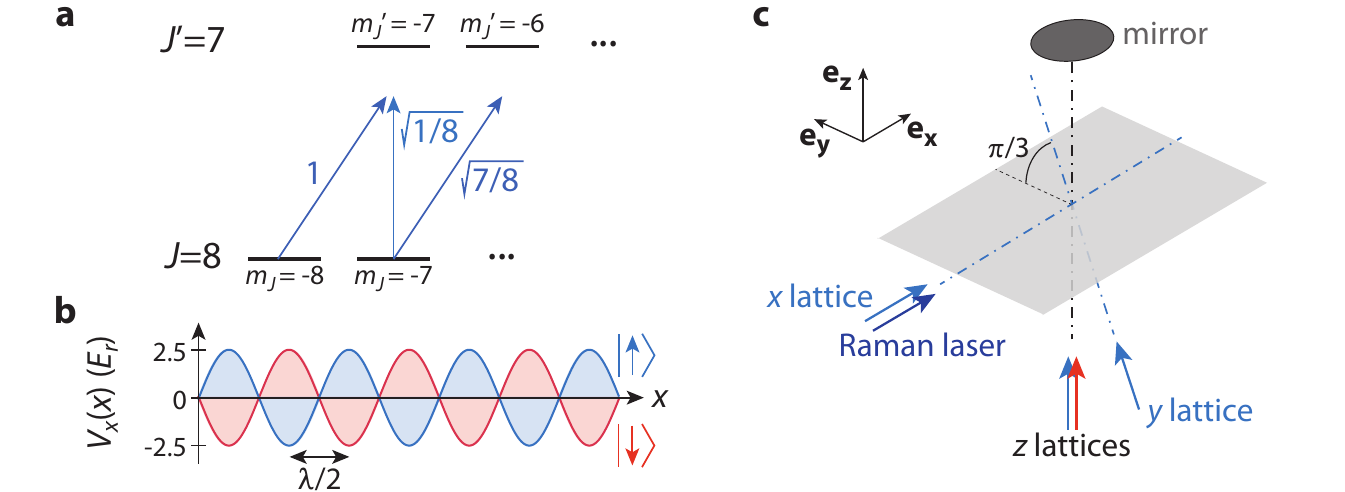}
\vspace{-5mm} \caption{
\textbf{a.} Scheme of the electronic states used for creating spin-dependent optical lattices with a $J\rightarrow J'=J-1$ optical transition.
\textbf{b.} Lattice potential along $x$ obtained by interfering a standing wave along $z$ with a plane wave along $x$. The required spin-dependent lattice configuration is achieved for specific choices of polarization.
\textbf{c.} Geometry of the complete laser configuration generating the optical lattice potential and the laser-induced tunneling in the $x$-lattice. 
 \label{Fig_Spin-dependent_lattice}}
\end{figure}

The plane-dependent optical lattice along $y$ can be created  using the interference of a travelling wave propagating along $\mathbf u=\cos\alpha\,\ey+\sin\alpha\,\ez$ with a standing wave along $z$ (see figure \ref{Fig_Spin-dependent_lattice}c), creating a potential
\begin{eqnarray*}
V(\rr)&\propto&\left|E_ue^{ik\uy\cdot\rr}+E_z\cos(kz+\phi')\right|^2\\
&=&\left|E_z\right|^2\cos^2(kz+\phi')+\\
&&2|E_zE_u|\cos(kz+\phi')\cos(k(\cos\alpha\, y+\sin\alpha\, z))+|E_u|^2.
\end{eqnarray*}
The first term represents a simple optical lattice along $z$ that contributes to the superlattice potential. The second term consists in an optical lattice along $y$ of lattice spacing $d_y=\lambda/\cos\alpha$ whose amplitude depends on the plane positions $z_A$ and $z_B$. As the $z$ standing wave has a node on the retroreflection mirror, the phase $\phi'$ can be controlled with high precision by varying the laser frequency over typically a GHz range \cite{folling2007direct,Tarruell2012}. This allows one to choose the value of $\phi'$ so that the $y$ lattice amplitude vanishes at $z_B$, leaving the motion of molecules in the plane $B$ unaffected (see figure~\ref{Fig_Lattice}b,c).

Choosing the value $\alpha=\pi/3$ corresponds to a $y$ lattice spacing $d_y=2\,\lambda$. For such a value $d_y\gg a$ the amplitude for atom pair tunneling into different tubes is about $10^{-3}$ smaller than for pair tunneling into the same tube. This allows one to consider that the tubes are independent from each other. 

The same technique can be used to create the lattice potential along $x$. As the optical lattice wavelength is chosen close to a narrow transition, vector light shifts can be used to create the spin-dependent potential in the desired configuration  shown in figure \ref{Fig_Spin-dependent_lattice}b. We checked that using specific laser polarizations we can create the required spin-independent lattices along $y$ and $z$ and the spin-dependent lattice along $x$ (see \ref{appendix_polar}).

\subsubsection{Laser-induced tunneling}
In order to realize Kitaev's model atom tunneling in the $x$ lattice should be described by the Hamiltonian $-J\sum_i(-1)^i\hat c_i^\dagger\hat c_{i+1}$. The required alternation of sign of the tunnel amplitude can be achieved by inducing atom tunneling via Raman transitions in the following geometry (see figure~\ref{Fig_Spin-dependent_lattice}c): one of the Raman beams would be the $z$ lattice beam used for generating the $x$ and $y$ lattices, and the second beam would propagate along $x$, with a frequency detuning equal to the $\ket{\uparrow}\leftrightarrow\ket{\downarrow}$ transition frequency. The hopping matrix element between the sites $2i$ and $2i+1$ is then given by 
\[
J_{2i,2i\pm1}=\hbar\Omega_0\int\dd \rr\,\cos(kz+\phi_z)e^{ikx}W_{2i}(\rr)^*W_{2i\pm1}(\rr),
\]
where $W_{i}(\rr)=w_x(x-id)w_y(y)w_z(z)$ is the Wannier fonction of the site $i$ of the $x$ lattice and $\Omega_0$ is the Rabi frequency of the Raman transition in vacuum. An explicit calculation leads to the result
\[
J_{2i,2i\pm1}=\pm J,\quad J=\hbar\Omega_0\int\dd z\,\cos(kz+\phi_z)|w_z(z)|^2\int\dd x\,e^{ikx}w_x(x)^*w_x(x-d),
\]
which is the desired configuration, assuming the gauge is chosen so that $J$ is real.

\section{Conclusions and perspectives}
In this article we have discussed how to realize 1D topological superfluids with ultracold Fermi gases.  We described how to probe the Majorana fermions  located at the edges of the topological superfluid phase. Further developments could address the robustness of the Majorana degeneracy with respect to disorder \cite{akhmerov2011quantized,brouwer2011topological} and atomic interactions between neighbouring sites \cite{gangadharaiah2011majorana,stoudenmire2011interaction,sela2011majorana,lutchyn2011interacting}, which could be induced by the large dipole-dipole interactions for Dy or Er atoms\footnote{For two Dy atoms separated by $\lambda/2=265\,$nm the dipole-dipole interaction varies between $-0.03\,E_r$ and $0.016\,E_r$, depending on the magnetic field direction.}. On a longer term the manipulation of Majorana fermions would allow one to exhibit their non-Abelian statistics, which is the basic ingredient for developing topological quantum computation \cite{kitaev2003fault,freedman2003topological,das2005topologically}. The realization of braiding operations requires either implementing a network of coupled 1D tubes \cite{alicea2011non} or creating a two-dimensional $p_x+ip_y$ topological superfluid with Majorana fermions bound to half-quantum vortices \cite{read2000paired,das2005topologically,sarma2006proposal}. The experimental scheme presented in this article could be adapted to a 2D geometry to realize a lattice $p_x+ip_y$ topological superfluid.  

\ack 

We thank the members of the `Bose-Einstein condensates' and `Fermi gases' groups at LKB, the Quantum optics theory group at Innsbruck university, V. Gurarie, N. Cooper, and N. Goldman for helpful discussions. Laboratoire Kastler Brossel is unité mixte de recherche (UMR 8552) of the CNRS.

\appendix

\settocdepth{section}

\section[\hspace{2.3cm}Semi-classical approximation]{Semi-classical approximation\label{Appendix}}
In this appendix we provide details on the calculation of the proximity-induced gap using a semi-classical approximation. For replacing the mBEC coherent state by a classical field we use a unitary displacement of the Hamiltonian by the operator $\hat D_\mBEC=e^{-\sqrt{N}/2}\exp(\sqrt{N}(\hat\psi_{\mol}^\dagger-\hat\psi_{\mol}))$. For the calculation of $\hat D_\mBEC^\dagger\hat H_z\hat D_\mBEC$ we use the formula
\[
e^{\hat A}\hat Be^{-\hat A}=\hat B+\left[\hat A,\hat B\right]+\frac{1}{2!}\left[\hat A,\left[\hat A,\hat B\right]\right]+\frac{1}{3!}\left[\hat A,\left[\hat A,\left[\hat A,\hat B\right]\right]\right]+\ldots
\]
This leads to an effective Hamiltonian 
\begin{eqnarray*}
\hat H_\eff&=&-J_z\int\dd\rrho\dd\rrho'\;
f(\mathbf{\rrho}'-\mathbf{\rrho}')\left(\hat\psi_{A,\uparrow}^\dagger(\rrho)\hat\psi_{B,\uparrow}^{\phantom{\dagger}}(\rrho')+\hat\psi_{A,\downarrow}^\dagger(\rrho)\hat\psi_{B,\downarrow}^{\phantom{\dagger}}(\rrho')+\hc \right)\\
&&-J_z\int\dd\rrho_\uparrow\dd\rrho_\downarrow\;
g(\mathbf{\rrho}_\uparrow-\mathbf{\rrho}_\uparrow)\left(\hat\psi_{B,\uparrow}^\dagger(\rrho_\uparrow)\hat\psi_{A,\downarrow}^\dagger(\rrho_\downarrow)+\hat\psi_{A,\uparrow}^\dagger(\rrho_\uparrow)\hat\psi_{B,\downarrow}^\dagger(\rrho_\downarrow)+\hc \right),
\end{eqnarray*}
where the first terms describes atom tunneling between the planes $A$ and $B$ and the second terms corresponds to the creation/annihilation of atom pairs from the mBEC.

The functions $f(\rho)$ and $g(\rho)$ are given by
\begin{eqnarray*}
f(\rho)&=&\delta(\rho)+\sum_{n=1}^\infty\frac{(-1)^n}{(2n)!}\left(\frac{k_Fa}{2\pi}\right)^{2n}K_0^{*2n}(\rho),\\
g(\rho)&=&\sum_{n=0}^\infty\frac{(-1)^{n+1}}{(2n+1)!}\left(\frac{k_Fa}{2\pi}\right)^{2n+1}K_0^{*2n+1}(\rho),
\end{eqnarray*}
where $K_0^{*i}$ refers  to the function $K_0$ convolved with itself $i$ times. 

In the main text we only kept for both functions the first term of the series. We checked that for the value $E_F=E_b/4$ used for the calculations this assumption is justified (error on $g(\rho)$ smaller than $10\%$).

\section[\hspace{2.3cm}Polarization of the optical lattices]{Polarization of the optical lattices\label{appendix_polar}}
We provide in table \ref{Table_Polar} a possible choice for the polarization of the lasers used for the spin-independent lattices along  $z$ and $y$, and the spin-dependent lattice along $x$.

 \begin{table}[b]
\caption{\label{Table_Polar}Table of numerical values for the magnetic field direction and laser directions/polarizations leading to spin-independent lattices along  $z$ and $y$, and a spin-dependent lattice along $x$ in the required configuration.} 

\begin{indented}
\lineup
\item[]
\hspace{-2.3cm}\begin{tabular}{@{}*{3}{l}}
\br                              
&Direction & Polarization\cr 
\mr
$B$ field & $\mathbf{u}_B=\cos\beta\,\ey+\sin\beta\,\ez$, $\beta=0.43\,$rad \cr
 $z$ standing wave & $\ez$ & $\cos\theta_z\,\ex+i\sin\theta_z\,\ey,$ $\theta_z=0.86\,$rad \cr 
 $x$ travelling wave & $\ex$ & $\cos\theta_x\,\ey+\sin\theta_x\,\ez,$ $\theta_x=0.53\,$rad \cr 
  $y$ travelling wave & $\mathbf{u}_y=\cos\alpha\,\ey+\sin\alpha\,\ez$, $\alpha=\pi/3$ & $\cos\theta_y\,\ex\times\uy+\sin\theta_y\,\ex,$ $\theta_y=1.05\,$rad\cr 
\br
\end{tabular}
\end{indented}
\end{table}

\section*{References}

\bibliographystyle{unsrt}
\bibliography{biblio}

\end{document}